\documentclass{ifacmtg}
\usepackage{harvard}
\usepackage{graphicx,amssymb}
\usepackage{bm}
\usepackage{epsfig}
\begin{document}
\runauthor{T. Benzekri, C. Chandre, X. Leoncini, R. Lima, M. Vittot, A. Goullet and N. Aubry  }

\begin{frontmatter}
\title{Control of chaotic advection}
\author[Paestum]{T. Benzekri\thanksref{Someone}}
\author[Paestum]{C. Chandre}
\author[Paestum]{X. Leoncini}
\author[Paestum]{R. Lima}
\author[Paestum]{M. Vittot}
\address[Paestum]{Centre de Physique Th\'eorique, CNRS Luminy, Case 907, F-13288 Marseille cedex 9, France}
\author[Baiae]{A. Goullet}
\author[Baiae]{N. Aubry}
\address[Baiae]{Department of Mechanical Engineering \\
New Jersey Institute of Technology, Newark, New Jersey 07102}
\thanks[Someone]{benzekri@cpt.univ-mrs.fr}
\begin{abstract}
A method of chaos reduction for Hamiltonian systems is applied to control chaotic advection. By adding a small and simple term to the stream function of the system, the construction of invariant tori has a stabilization effect in the sense that these tori act as barriers to diffusion in phase space and the controlled Hamiltonian system exhibits a more regular behaviour.
\end{abstract}
\end{frontmatter}

\section{Introduction}

Chaotic advection (or Lagrangian chaos) was introduced by Aref (1988) to qualify a motion in which it is possible to generate chaotic trajectories even if the flow is laminar. Such phenomena are observed in a wide range of physical systems  (Ottino (1989)) and has fundamental applications for instance in geophysical flow (Behringer, et al. (1991); Brown and Smith (1991)). 
Chaotic advection was mostly studied for two dimensional unsteady flow 
(Solomon and Gollub (1988); Solomon, et al. (2003); Camassa and Wiggins (1991)). 
The advantage of these studies is that it uses the theory of dynamical systems and in particular the Hamiltonian theory for incompressible flows; the phase space being the physical space in this case. For these motions, experiments can be implemented in a laboratory. \\
Even though chaos is
sometimes  preferable in mixing problems (Benzekri, et al.(2006)), we usually want to suppress it for instance in chaotic advection.
Recently, a method to control chaotic diffusion in Hamiltonian dynamical systems was proposed by (Vittot, et al. (2005); Chandre, et al.(2005)). 
It was shown that it is possible to prevent chaotic diffusion by adding a small term to the Hamiltonian. This technique was used to a model describing anomalous transport in magnetized plasmas, and applied experimentally on a beam of electrons produced by a long Travelling Wave Tube (Chandre, et al.(2005)).\\
Our goal in this paper is to use this method of  control of Hamiltonian chaos for the problem of chaotic advection in a two dimensional time periodic flow.  We apply this method within the framework of an experiment using magneto-hydrodynamic technique which shows that particle trajectories in a time periodic flow are chaotic.  More precisely, the experiment consists in an electric current passing through a thin layer of an electrolytic solution with a free surface. The dynamics of passive particles of a time periodic two dimensional and incompressible flow is Hamiltonian. The Hamiltonian modeling the flow was derived ad hoc in (Solomon and Gollub (1988); Solomon, et al. (2003)). A comparison was made with the experiment to validate the model. The equation of motion of these passive particles comes from a Hamiltonian of $1.5$ degrees of freedom that we study in the limit of weak amplitude oscillations.\\
 The model proposed to describe this phenomena is 
based on the following streamfunction:
\begin{eqnarray}
\label{1}
\Psi_{\epsilon}(x,y,t)=\alpha \sin(x + \epsilon  \sin  t) \sin y,	
\end{eqnarray}
where 
$\alpha$ is the maximal vertical velocity in the flow, 
$\epsilon$  is the amplitude of the lateral oscillations of the velocity field. 
The current interacts with an alternative magnetic field produced by magnets below the fluid. A chain of vortices are then observed. Time periodic  dependence is imposed externally with a plunger that oscillates slowly up and down, displacing the flow laterally,  i.e., in the direction perpendicular to the roll axes, giving rise to chaotic advection.  
The same phenomenon is observed in  Rayleigh-B\'enard convection.
For the steady state, rolls are formed periodically. It has been shown that by imposing a sinusoidal flow, chaotic advection occurs.\\
Under the perturbation, i.e. for  $\epsilon\neq0$, the vertical heteroclinic connection breaks down and the stable and unstable manifold intersect transversely thus generating chaotic advection of passive particles.\\  
In this method, 
 we seek for one of the simplest perturbations to create barriers around the broken separatrix
 of the integrable case. We choose a perturbation depending only on the position variables.\\
We will use the stream function (\ref{1}) as the starting point to control or reduced the chaotic advection.

\section{Local control method}
\label{sec:control local} 
The local control method has been extensively described in (Vittot, et al. (2005))  where the corresponding rigorous  results were proved. We summerize here the results of this paper.
For a Hamiltonian system  with $L$ degrees of freedom, the perturbed Hamiltonian is 
$$
H({\bf A},{\bm\theta})={\bm \omega}\cdot {\bf A}+ V({\bf A},{\bm\theta}),$$ 
where $({\bf A},{\bm\theta})\in {\mathbb R}^L\times {\mathbb T}^L$ are action-angle like variables and $\bm\omega$ is a non-resonant vector of ${\mathbb R}^L$. 
Without loss of generality, let us consider a region near ${\bf A}={\bf 0}$ (by translation), since the Hamiltonian is nearly integrable, the perturbation $V$ has constant and linear parts in actions of order $\varepsilon$, i.e.\ 
\begin{equation}
\label{eqn:e4V}
V({\bf A},{\bm\theta})=\varepsilon v({\bm\theta})+\varepsilon {\bf w}({\bm\theta})\cdot {\bf A}+Q({\bf A},{\bm\theta}),
\end{equation} 
where $Q$ is of order $O(\Vert {\bf A}\Vert ^2)$. Note that for $\varepsilon=0$, the Hamiltonian $H$ has an invariant torus with frequency vector ${\bm\omega}$ at ${\bf A}={\bf 0}$ for any $Q$ not necessarily small.
The controlled Hamiltonian  is then constructed as:
\begin{equation}
\label{eqn:gene}
H_c({\bf A},{\bm\theta})={\bm \omega}\cdot {\bf A}+ V({\bf A},{\bm\theta})+  f({\bm \theta}).
\end{equation}
 In this situation,  the control term $f$  only depends on the angle variables and is given by
\begin{equation}
\label{eqn:exf}
f({\bm\theta})=V({\bf 0},{\bm\theta})-V\left( -\Gamma \partial_{\bm\theta} V({\bf 0},{\bm\theta}),{\bm\theta}\right),
\end{equation}
where $\Gamma$ is a linear operator defined as a pseudo-inverse of ${\bm\omega}\cdot \partial_{\bm\theta}$, i.e.\ acting on $V=\sum_{{\bf k}}V_{{\bf k}} {\mathrm e}^{i{\bm\theta}\cdot{\bf k}}$ as
\begin{equation}
\label{gam}
\Gamma V=\sum_{{\bm\omega}\cdot{\bf k}\not= 0} \frac{V_{{\bf k}}}{i{\bm\omega}\cdot{\bf k}} {\mathrm e}^{i{\bm\theta}\cdot{\bf k}}.
\end{equation}
Note that $f$ is of order $\varepsilon^2$. This can be seen from Eq.~(\ref{eqn:e4V}) since $f$ can be rewritten as
$$
f({\bm\theta})=\varepsilon^2 {\bf w}({\bm\theta})\cdot \Gamma \partial_{\bm\theta} v-Q\left(  -\varepsilon \Gamma \partial_{\bm\theta} v, {\bm\theta} \right),
$$
and $Q$ is quadratic in the actions.
For any perturbation $V$, Hamiltonian~(\ref{eqn:gene}) has an invariant torus with frequency vector close to ${\bm\omega}$.
The equation of the torus which is constructed by the control is
\begin{equation}
\label{eqn:eto}
	{\bf A}=-\Gamma \partial_{\bm\theta} V({\bf 0},{\bm\theta}),
\end{equation}
which is of order $\varepsilon$ since $ V({\bf 0},{\bm\theta})$ is of order $\varepsilon$.

In the next section, we will see that the amplitude of the control term is small compared with the perturbation. 

\subsection{Application to the suppression of chaotic advection}

In this section, we apply the method previously summerized to reduce chaotic transport of passive tracers. With this method, we create isolated barriers of transport. In particular, a barrier created between two convection rolls allows one to reduce the diffusion across these rolls. Let us first give some results in the integrable case.\\
For $\epsilon=0$, the Hamiltonian is integrable, the trajectories of advected particles coincide with streamlines. The phase space, which is here the physical space,  is characterized by a chain of rolls with separatrices localized at $x=m \pi$, with $m  \in \mathbb Z$. The fluid is limited by two invariant surfaces $y=\pi$ and $y=0$ corresponding to the top and bottom roll boundaries.\\
In the integrable case, as mentionned in (Camassa and Wiggins (1991)), the flow is characterized by hyperbolic fixed points along the two invariant surfaces $y=\pi$ and $y=0$ and localized at $x=m \pi$, $m$  $\in \mathbb Z$. These points are joined by a vertical heteroclinic connection for which the stable and unstable manifolds coincide.   \\
In order to built a barrier, we select a surface  which is located around $x=0$.  However we could also create a barrier at $x=\pi$ (mod $2\pi$)\\
We map the time-dependent stream function $\Psi_{\epsilon}$ given by Eq. ~(\ref{1}) into an  autonomous Hamiltonian  written as $H(x,y,E,\tau)=E+\Psi_{\epsilon}(x,y,\tau)$. This corresponds to an extension of phase space
 $(x,y,E,\tau)$, where ${\bf A}=(x,E)$  and 
$\boldsymbol{\theta}=(y,\tau)$ are momenta and positions. We assume that at $t=0$, $\tau(0)=0$ for $H$.
Therefore the equation of motion of $H$ are the same as the ones for $\Psi_{\epsilon}$ since $\tau=t$.\\
We rewrite the autonomous Hamiltonian in the form:
\begin{eqnarray}
\label{34}
H(x,y,E,\tau)&= &E+ \alpha \sin(x +  \epsilon \sin \tau) \sin y,\nonumber \\
&=& H_0(x,E)+ \Psi_{\epsilon}(x,y,\tau),
\end{eqnarray}
and  develop the stream function  in Taylor series around $x=0$
\begin{eqnarray}
H(x,y,E,\tau)&=& E+ \epsilon v(y,\tau)+ \epsilon w(x,y,\tau) x \nonumber \\
&+& Q(x,E,y,\tau),
\end{eqnarray}
where $H_0(x,E)=E$ is the integrable part, $\epsilon v(y,\tau)=\Psi_{\epsilon}(0,y,\tau)$ and
\begin{eqnarray}
\label{35}
\epsilon w(x,y,\tau)&=&	\partial_x \Psi_{\epsilon}(0,y,\tau),\nonumber	\\
Q(x,E,y,\tau)&=&\sum_{l=1}^\infty \frac{1}{(l+1)!} \left[\partial_x^{l+1} \Psi_{\epsilon} (0,y,\tau)\right] x^{l+1}.	\nonumber
\end{eqnarray}
The frequency vector $\boldsymbol{\omega}$ is given by\\ 
$\boldsymbol{\omega} = \left({\partial H_0}/{\partial x}, {\partial H_0}/{\partial E}\right)=\left(0,1\right)$ and is  resonant.\\

In order to compute the operator $\Gamma$ given by Eq.(\ref{gam}), we rewrite $\Psi_{\epsilon}$ by  expanding $\sin (\epsilon \sin t)$ and $\cos (\epsilon \sin t)$ as  series of Bessel functions of first kind:
\begin{eqnarray}
\label{9}
\Psi_{\epsilon}(x,y,t)&=&
\alpha\sin y \sin x (\mathcal{J}_{0}^{\epsilon}+ 2 \sum_{n \geq 1} \mathcal{J}_{2n}^{ \epsilon} \cos 2 n  t )\nonumber \\
 &+& 2 \alpha\sin y \cos x ( \sum_{n \geq 1} \mathcal{J}_{2n+1}^{\epsilon} \sin (2n+1)  t ),\nonumber
\end{eqnarray}
where $\mathcal{J}_{l}^{\epsilon}=\mathcal{J}_{l}(\epsilon)$ for  $l \in \mathbb N$, and $\mathcal{J}_{l}$ are Bessel functions of the first kind.\\
The control term is given by Eq.(\ref{eqn:exf}):
\begin{eqnarray}
\label{37}
f(y,t)= \Psi_{\epsilon} (0,y,t)-\Psi_{\epsilon}( -\Gamma \partial_{y} \Psi_{\epsilon},y,t).\nonumber \\
\end{eqnarray}
Since $\Psi_{\epsilon}$ does not depend on $E$, we only have to compute  $\Gamma \partial_{y} \Psi_{\epsilon}(0,y,t)$ and then
\begin{eqnarray}
\label{30}
f(y,t) &=&\alpha\sin( \epsilon \sin t )	\sin y \nonumber \\
&& - \alpha\sin \left[-\alpha\cos y {C}_{\epsilon}(t)+\epsilon \sin t\right]\sin y,
\end{eqnarray}
where
\begin{eqnarray}
\label{serbessl}
C_{\epsilon}(t)&=&\Gamma \sin(\epsilon \sin t),\nonumber \\
&= &  \sum_{n \geq 0} \frac{-2}{2n+1}\mathcal{J}_{2n+1}^{\epsilon} \cos (2n+1)  t.
\end{eqnarray}

The equation of the invariant torus is given by Eq.~(\ref{eqn:eto})
\begin{eqnarray}
\label{bar}
x= \alpha \cos y  {C}_{\epsilon}(t).
\end{eqnarray}
The controlled stream function is given  by:

\begin{eqnarray}
\label{streamcont}
\Psi_{c}(x,y,t)&=&\alpha \sin y  \left( \sin(x+\epsilon\sin t)
+ \sin( \epsilon \sin t ) \right.\nonumber \\
&& -  \left. \sin \left[-\alpha \cos y {C}_{\epsilon}(t)+\epsilon \sin t\right] \right) .
\end{eqnarray}
 We notice that with a small modification of the stream function, provide that $\alpha$ and $\epsilon$ are such that $\mid \alpha \epsilon \mid < 1$  , there is an exact formula giving the equation of the torus. This invariant torus suppress the chaotic transport from roll to roll and then along the channel.\\
To illustrate numerically these results, we consider fixed values of the parameter $\alpha$ and $\epsilon$. Similar results hold for other values of the parameters $\alpha$ and $\epsilon$.

\begin{figure}
\begin{center}
\includegraphics[%
  width=3cm]{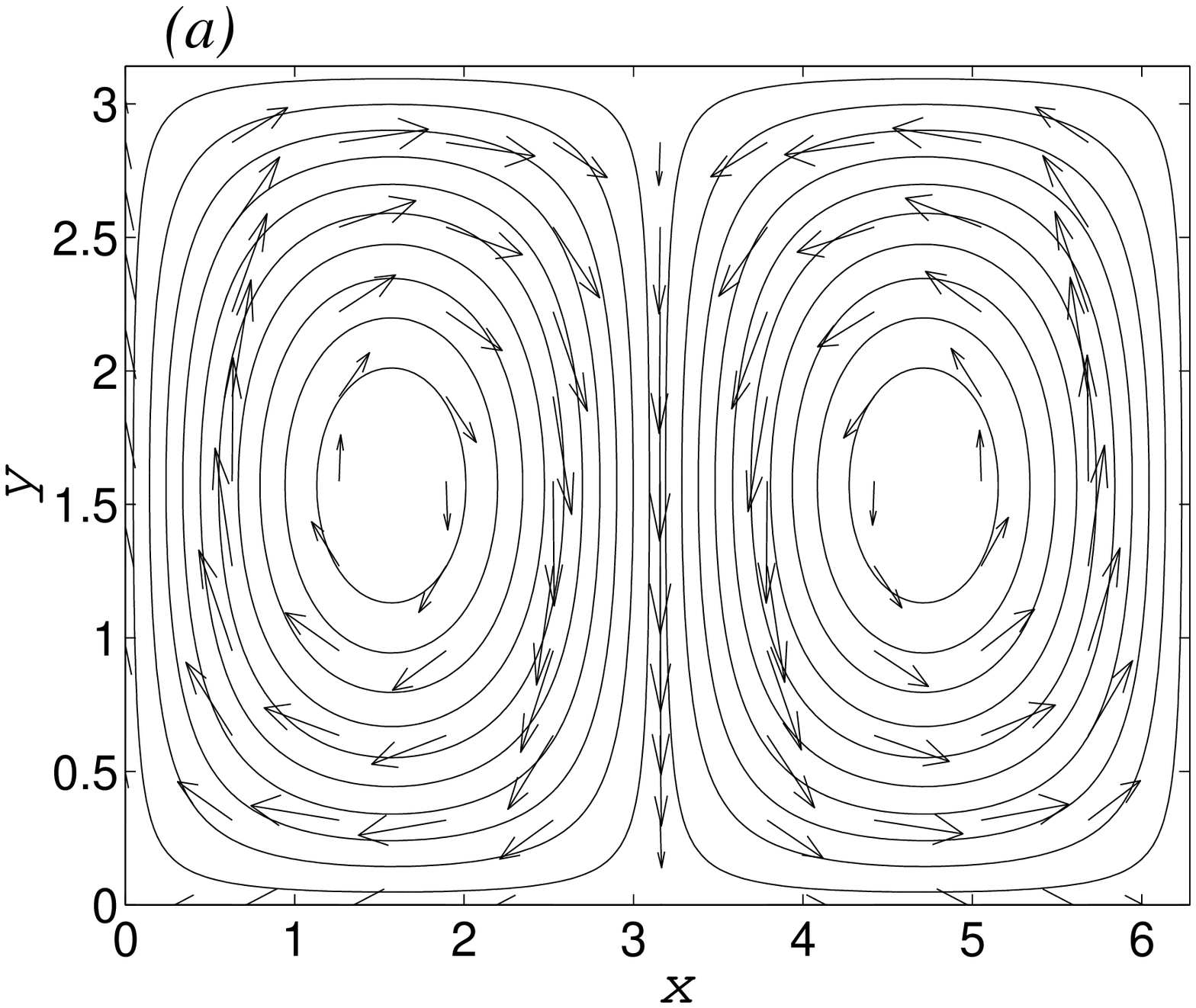}\includegraphics[%
  width=3cm]{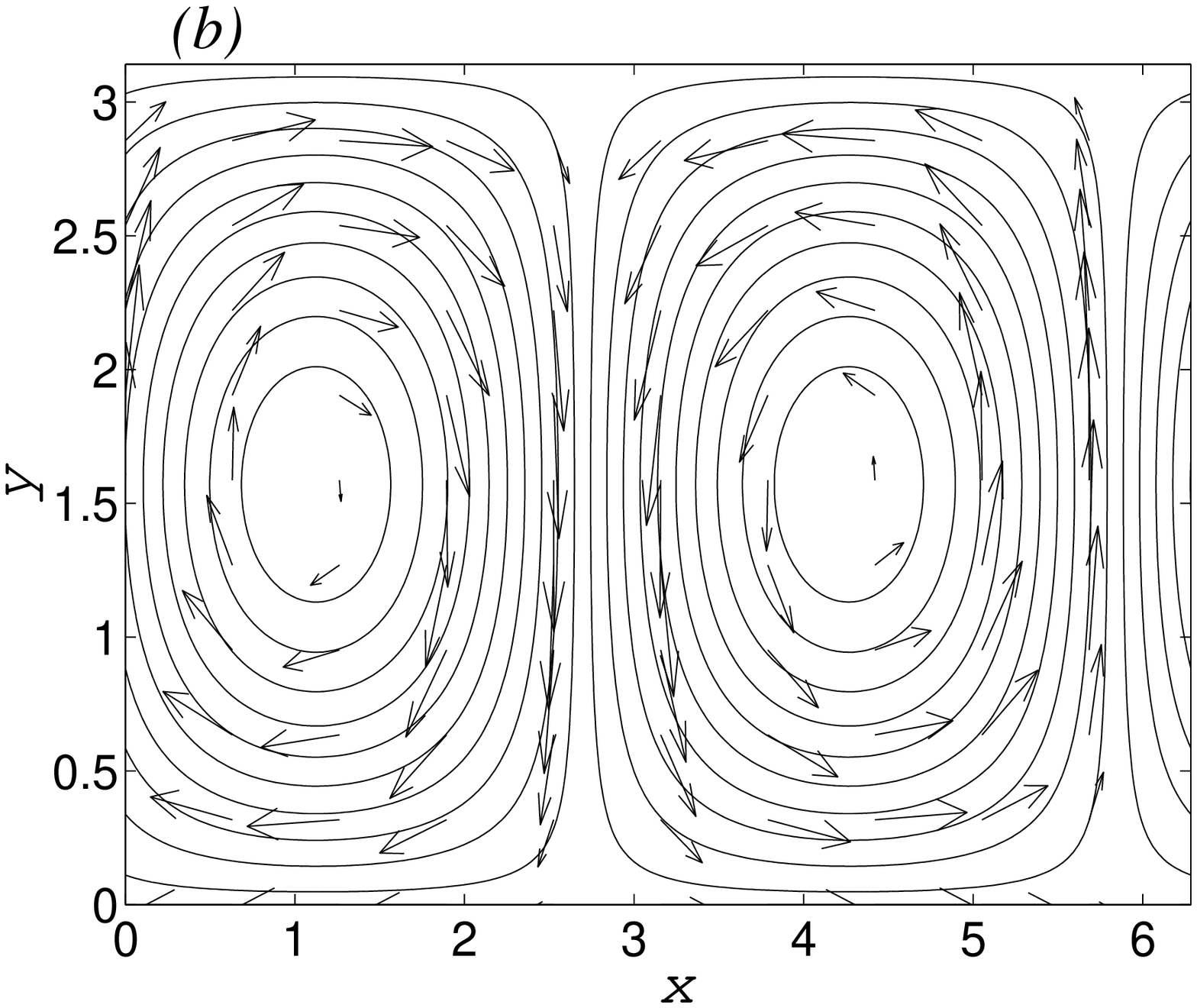}
 \end{center} 
  \begin{center}
\epsfig{file=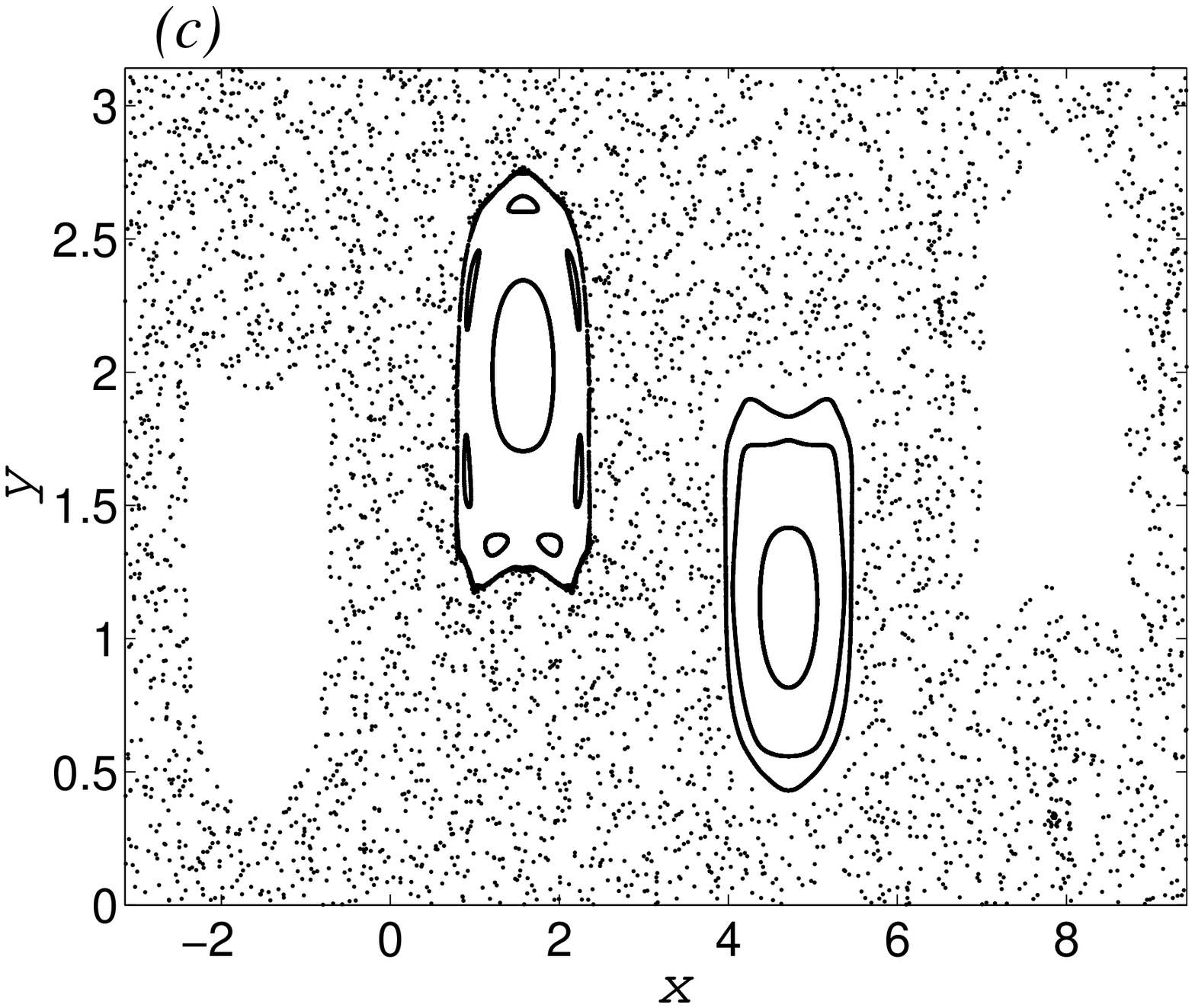,width=6.2cm,height=3cm}
 \end{center}
\caption{\label{Fig1} Streamlines at $(a)$ $t=0$ and $(b)$ $t=3\pi/4$, and $(c)$ Poincar\'{e} section 
 of the stream function (\ref{1}). The parameters are $\alpha=0.6$ and $\epsilon=0.63$.}
\end{figure} 

\begin{figure}
\begin{center}
\includegraphics[%
  width=3cm]{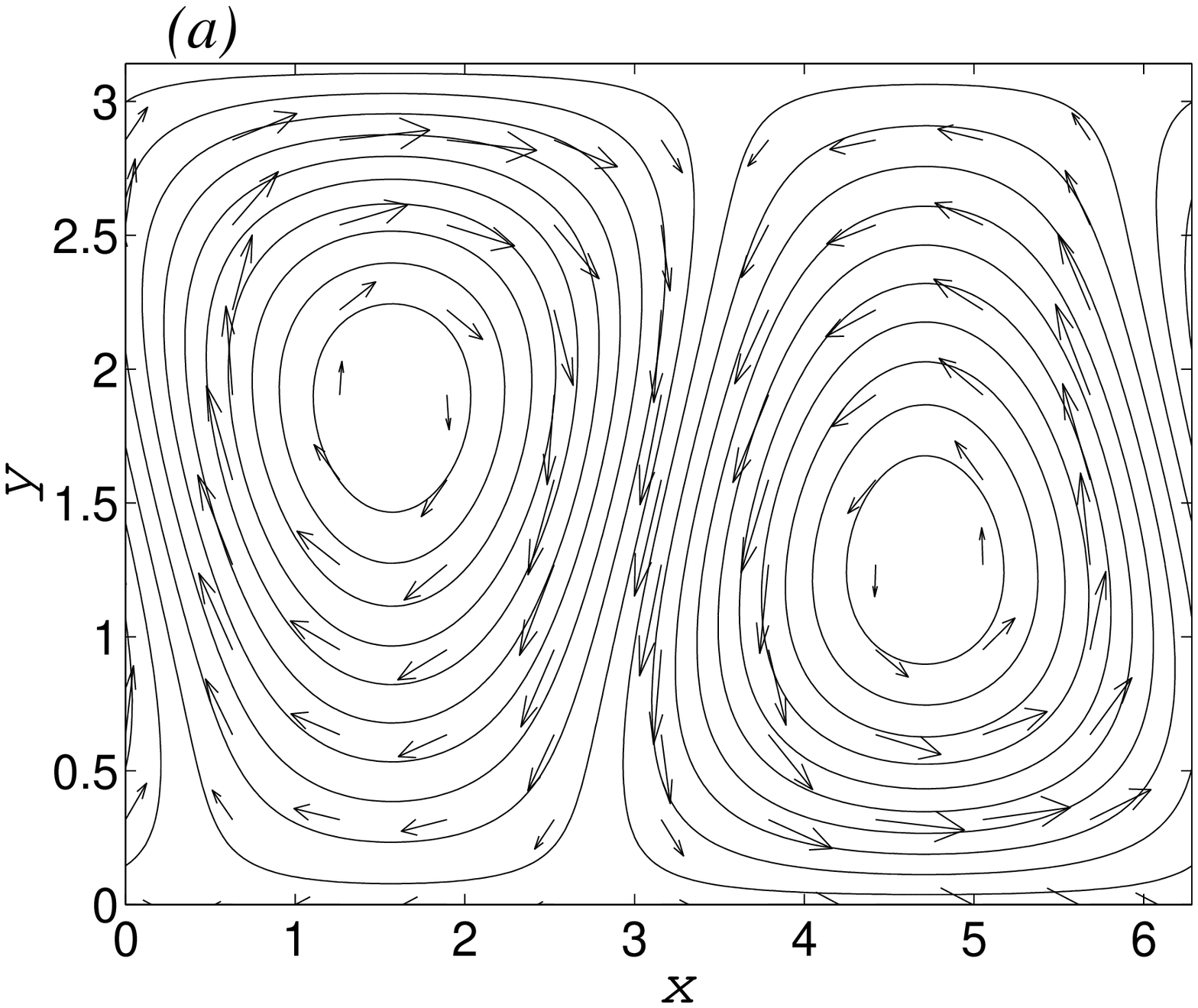}\includegraphics[%
  width=3cm]{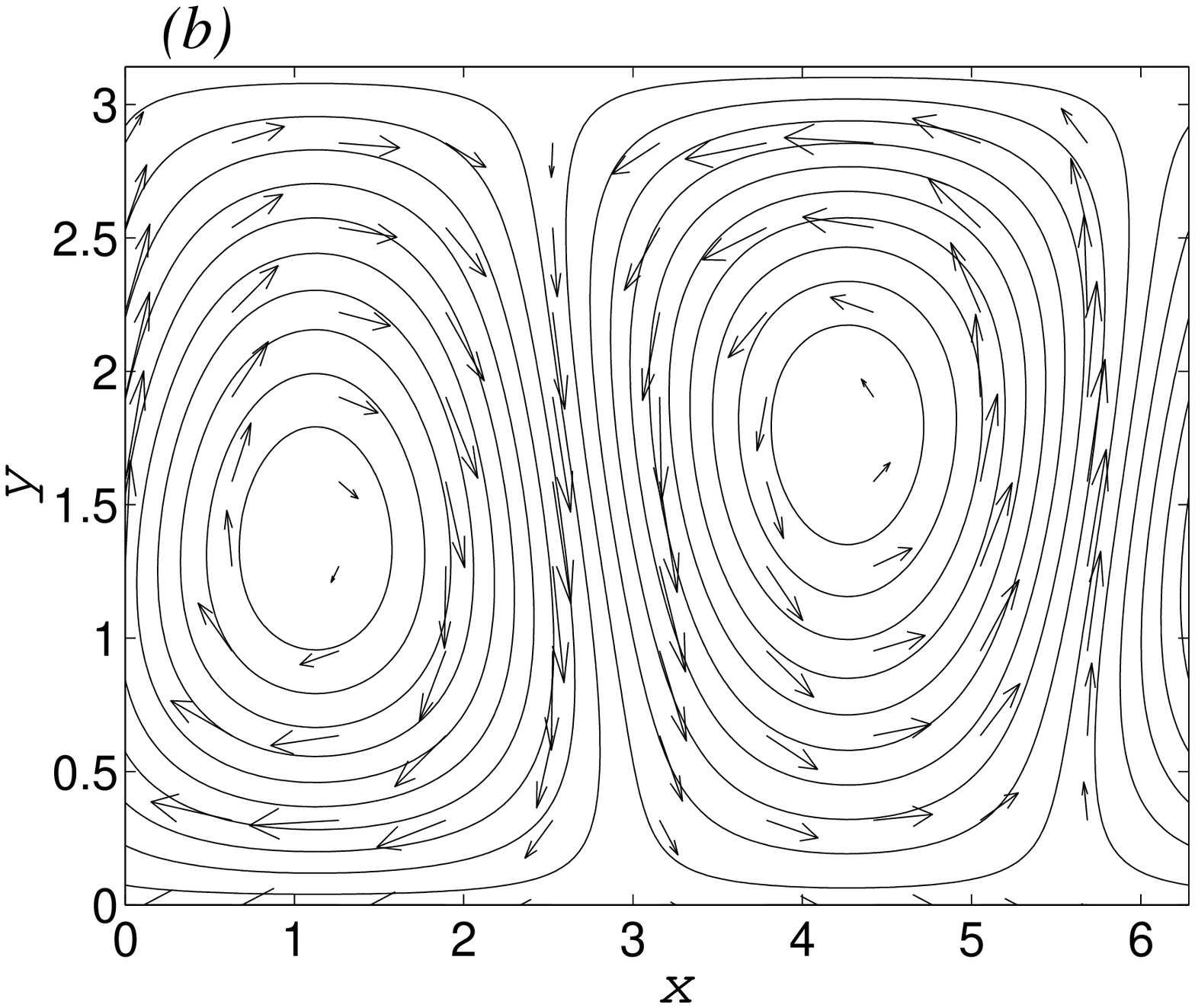}
 \end{center}  
  \begin{center}
  \epsfig{file=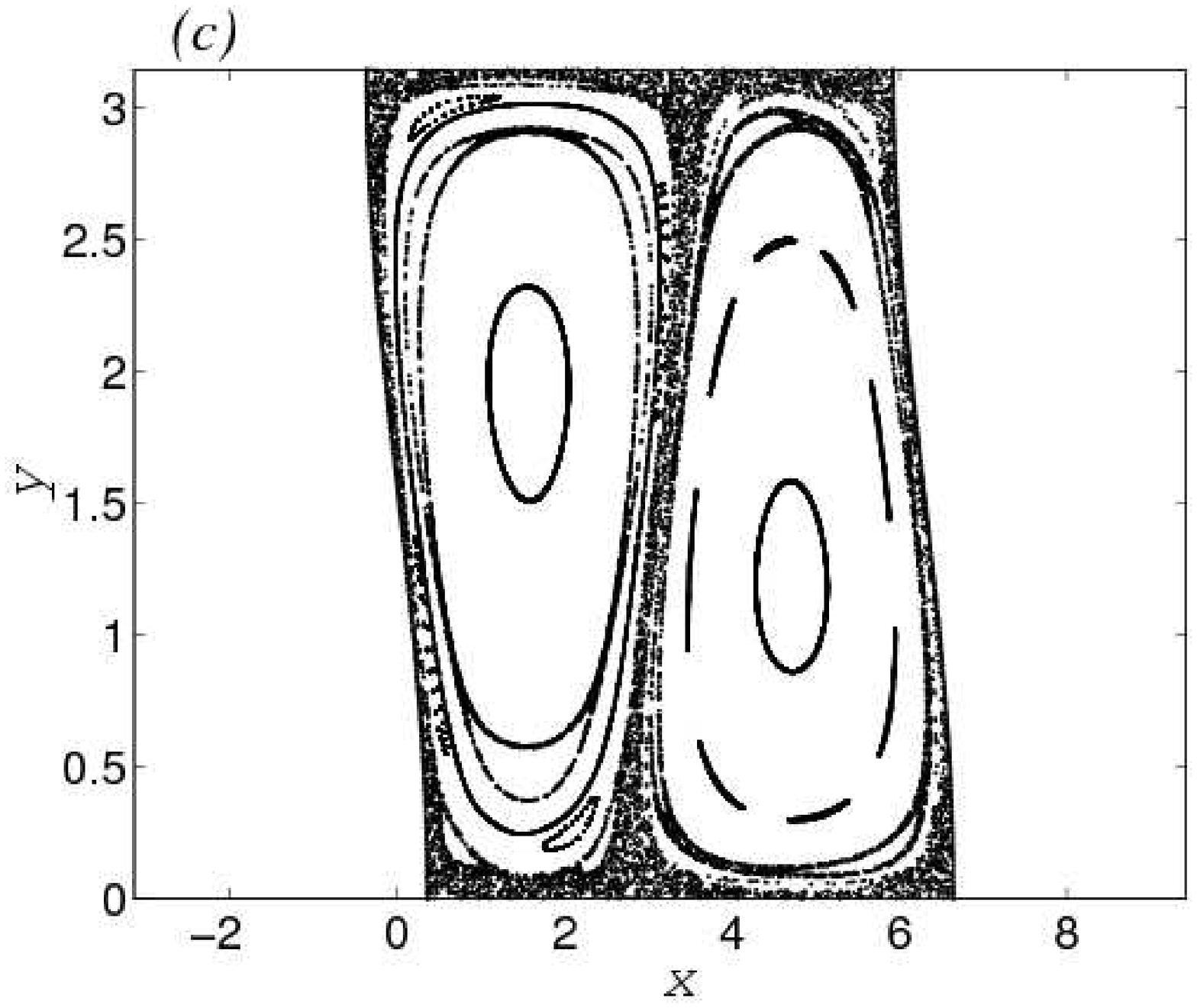,width=6.2cm,height=3cm}
 \end{center}  
 \caption{\label{Fig2} Streamlines at $(a)$ $t=0$ and $(b)$ $t=3\pi/4$, and $(c)$ Poincar\'{e} section 
 of the stream function (\ref{streamcont}). The parameters are $\alpha=0.6$ and $\epsilon=0.63$.}   
\end{figure}

Streamlines of the streamfunction (\ref{1}) are depicted on Fig.~\ref{Fig1}$(a)$ and $(b)$ at two different times $t=0$ and $t=3\pi/4$ respectively and for $\epsilon=0.63$ and $\alpha=0.6$. We observe 
  closed curves corresponding to lateral oscillations in the $x$-direction  with a periodic displacement of $-\epsilon\sin t$.
For the same values of $\epsilon$ and $\alpha$, a Poincar\'e section of the dynamics of the streamfunction (\ref{1}) for initial conditions without control are represented in Fig.~\ref{Fig1}$(c)$. It shows that the transport along the channel is greatly enhanced. 
The phase space is characterized by regular (quasiperiodic) trajectories and a chaotic region around and between the rolls.\\     
When we add the control term, the streamlines of the controlled stream function (\ref{streamcont}) are then slightly modified (non-uniformly in $y$) as it can be seen on Fig.~\ref{Fig2}$(a)$ and $(b)$ for $\epsilon=0.63$ and $\alpha=0.6$ at two different times $t=0$ and
$t=3\pi/4$ respectively. Moreover, the displacement of these rolls remains parallel to the $x$-axis as it is for the stream function given by Eq.~(\ref{1}). A Poincar\'{e} section of the dynamics of passive tracers in a flow
described by the stream function $\Psi_{c}$ given by Eq.~(\ref{streamcont})
is represented in Fig.~\ref{Fig2}$(c)$. We notice first  that there are invariant
surfaces which have been created around $x=0$ (mod $2\pi$) (bold
curves) which prevent the diffusion of  passive particles along the channel. Moreover we observe a regularisation of the dynamics whithin the cell bounded by the two barriers  $x=0$ and $\pi$ (mod $\pi$). \\

\begin{figure}
\includegraphics[%
  width=3cm]{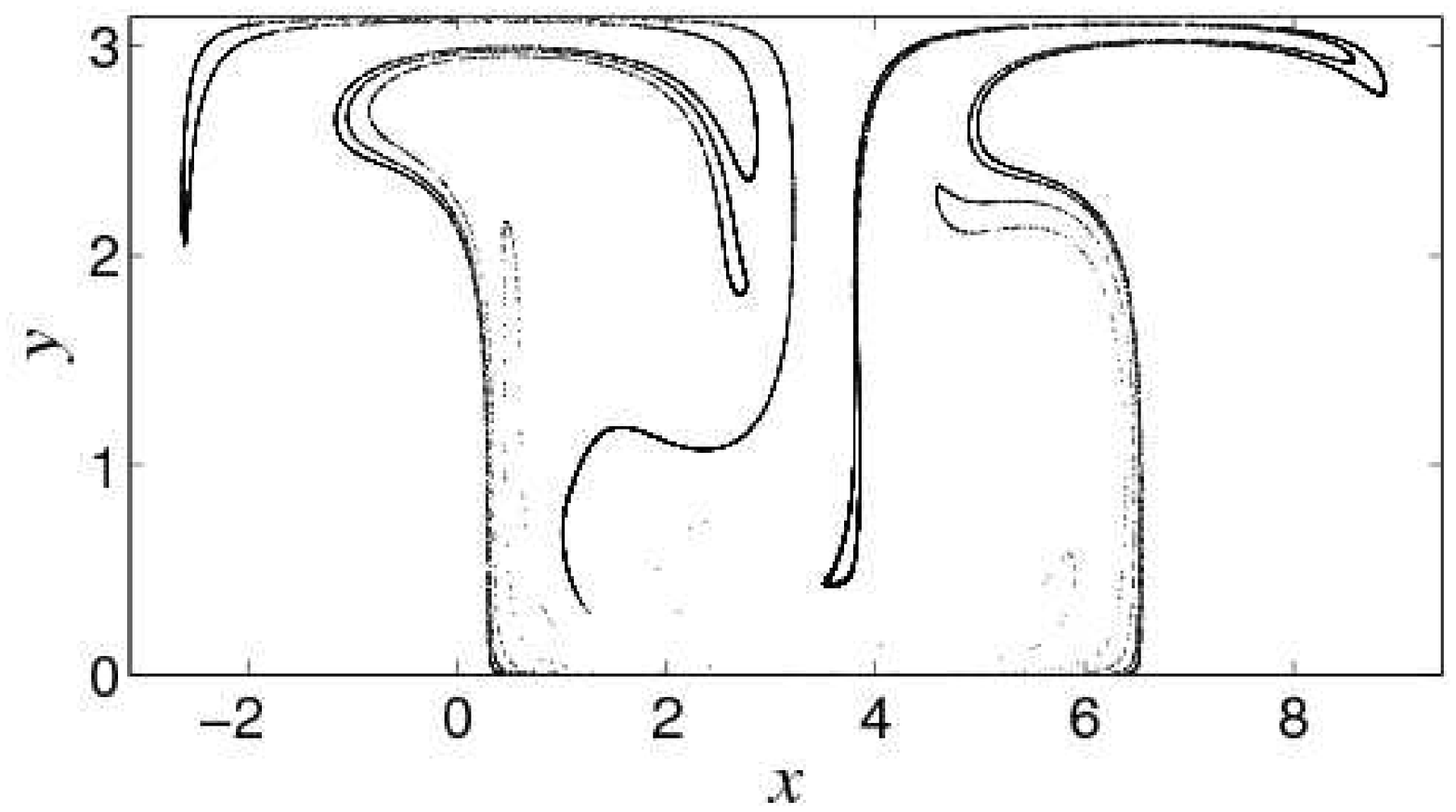}\includegraphics[%
  width=3cm]{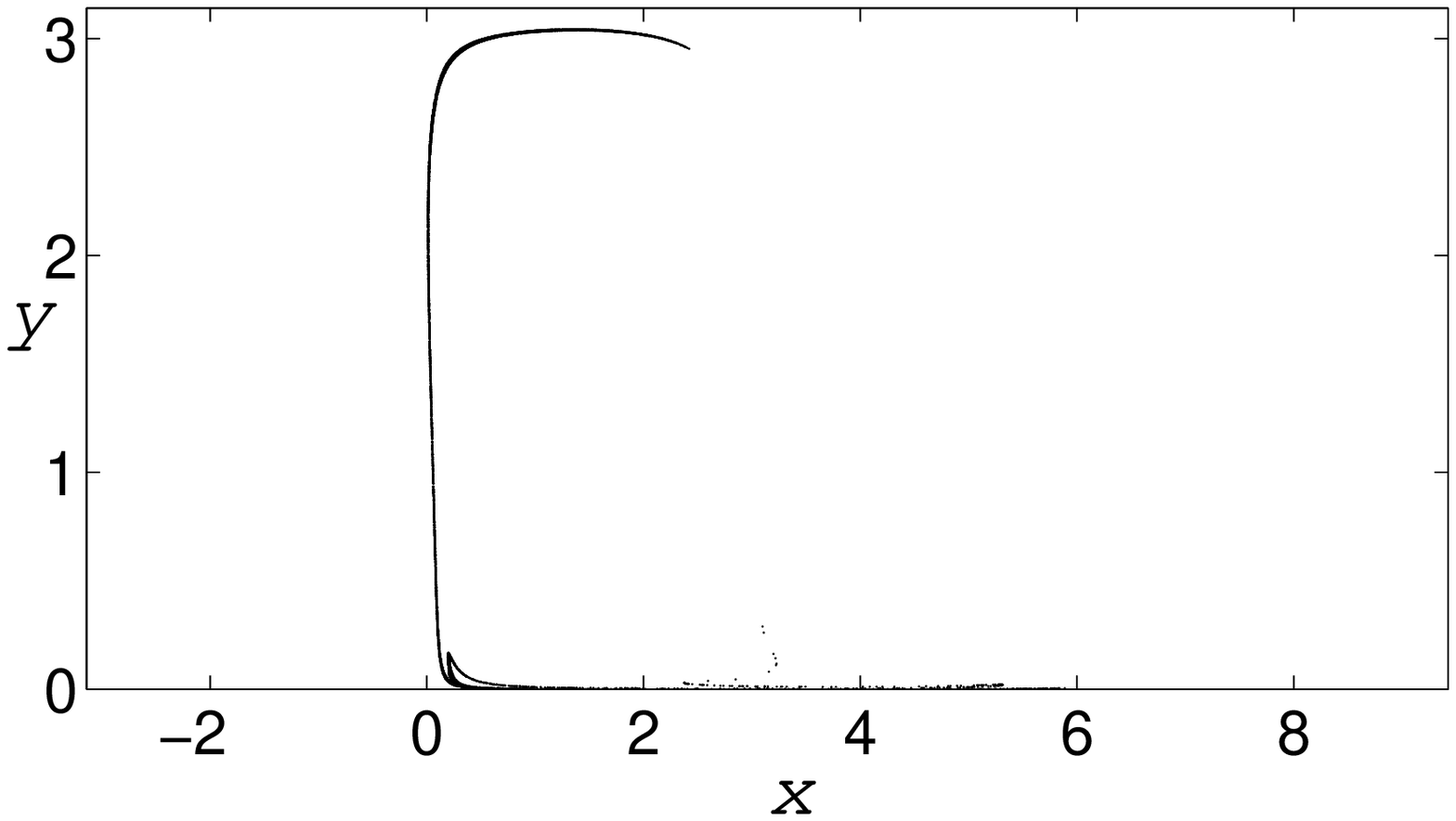}

\includegraphics[%
  width=3cm]{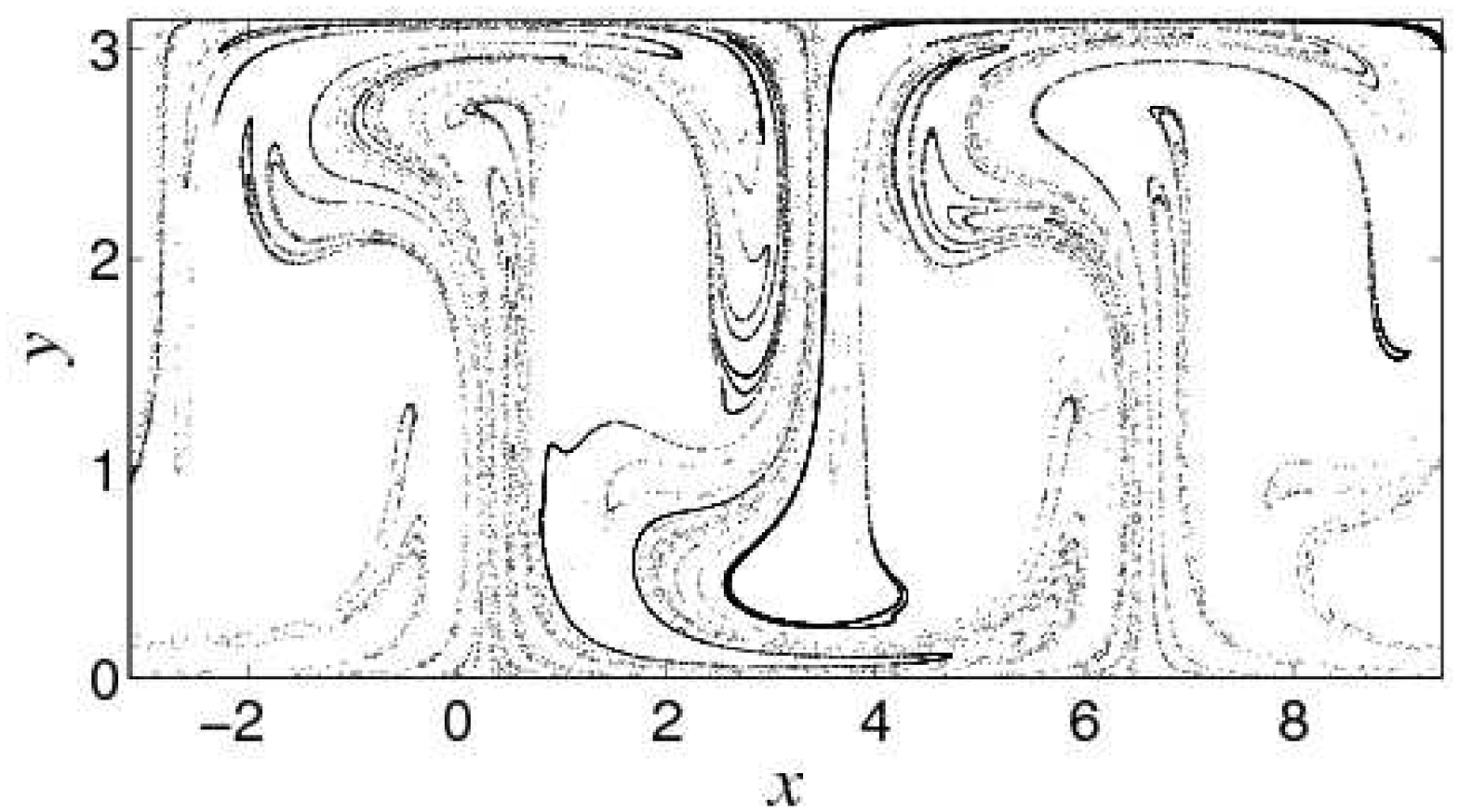}\includegraphics[%
  width=3cm]{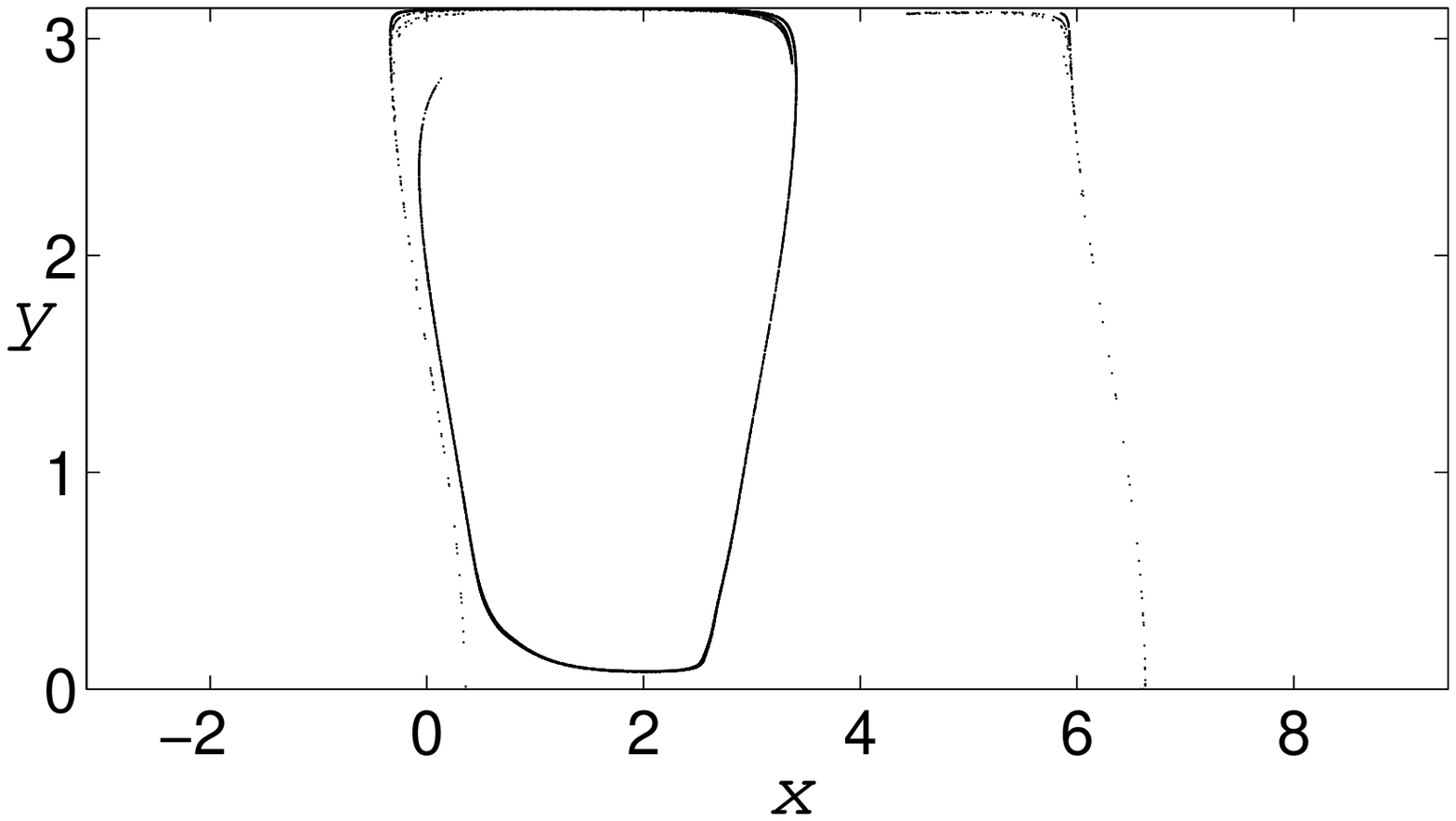}

\includegraphics[%
  width=3cm]{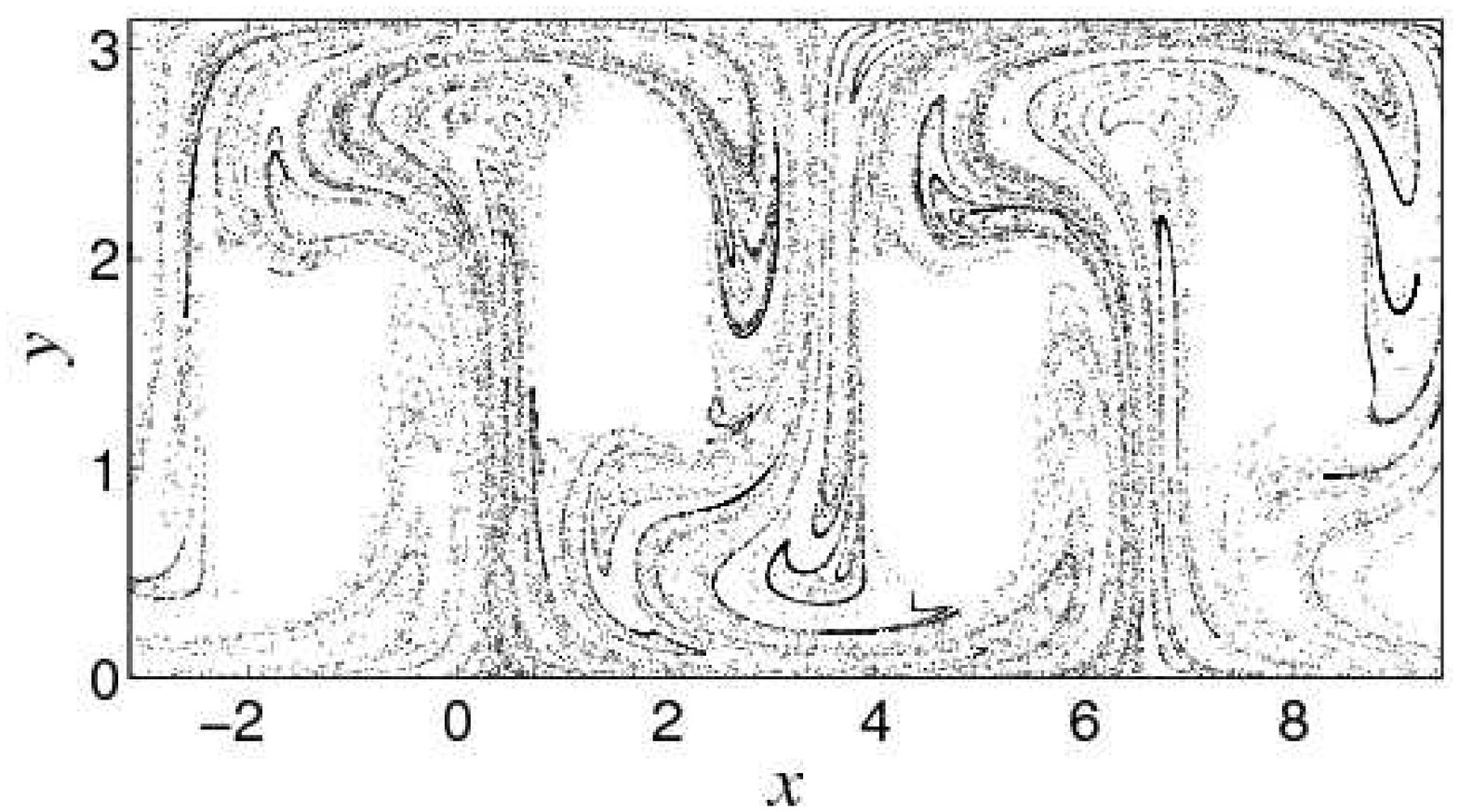}\includegraphics[%
  width=3cm]{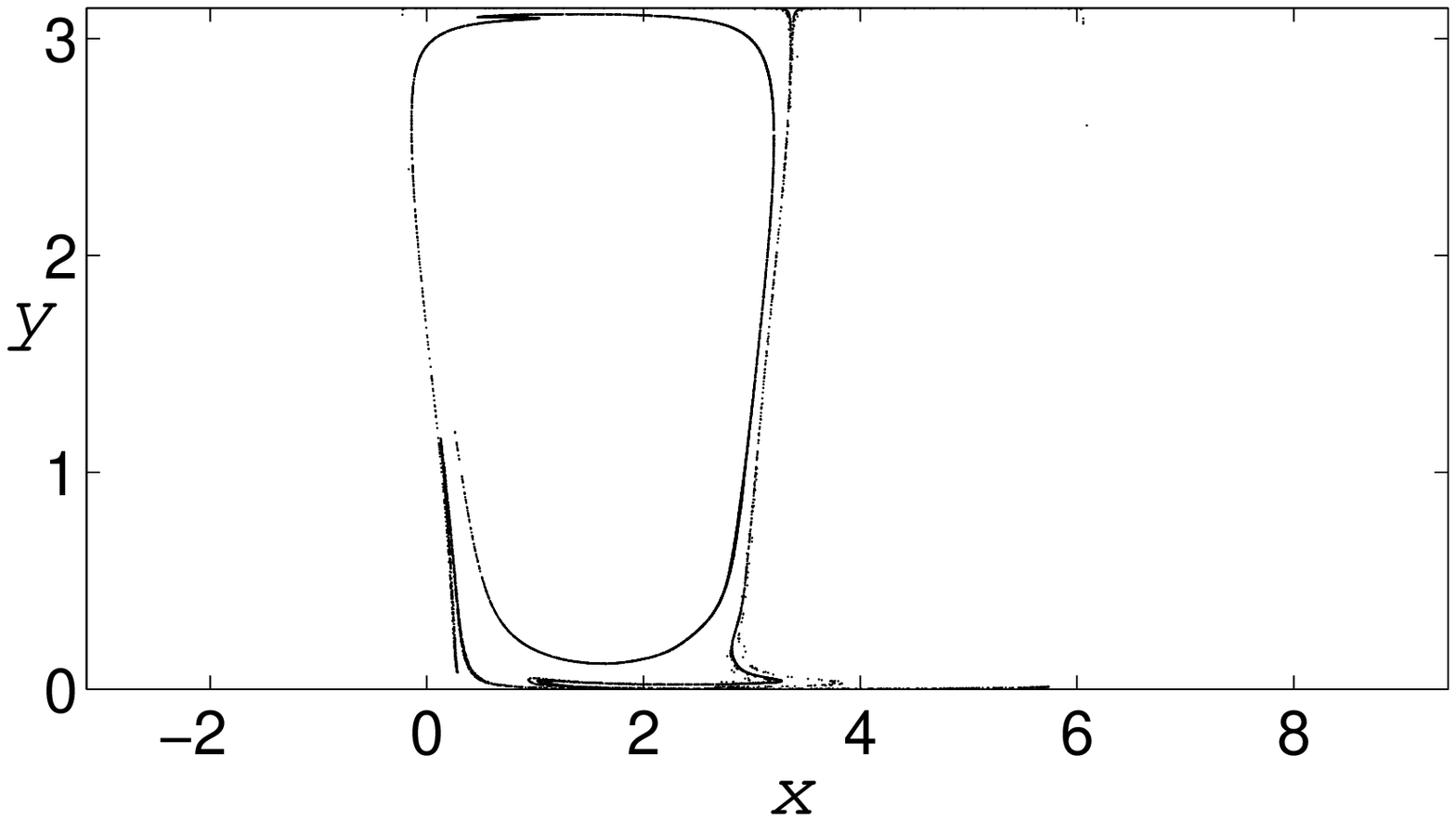}

\includegraphics[%
  width=3cm]{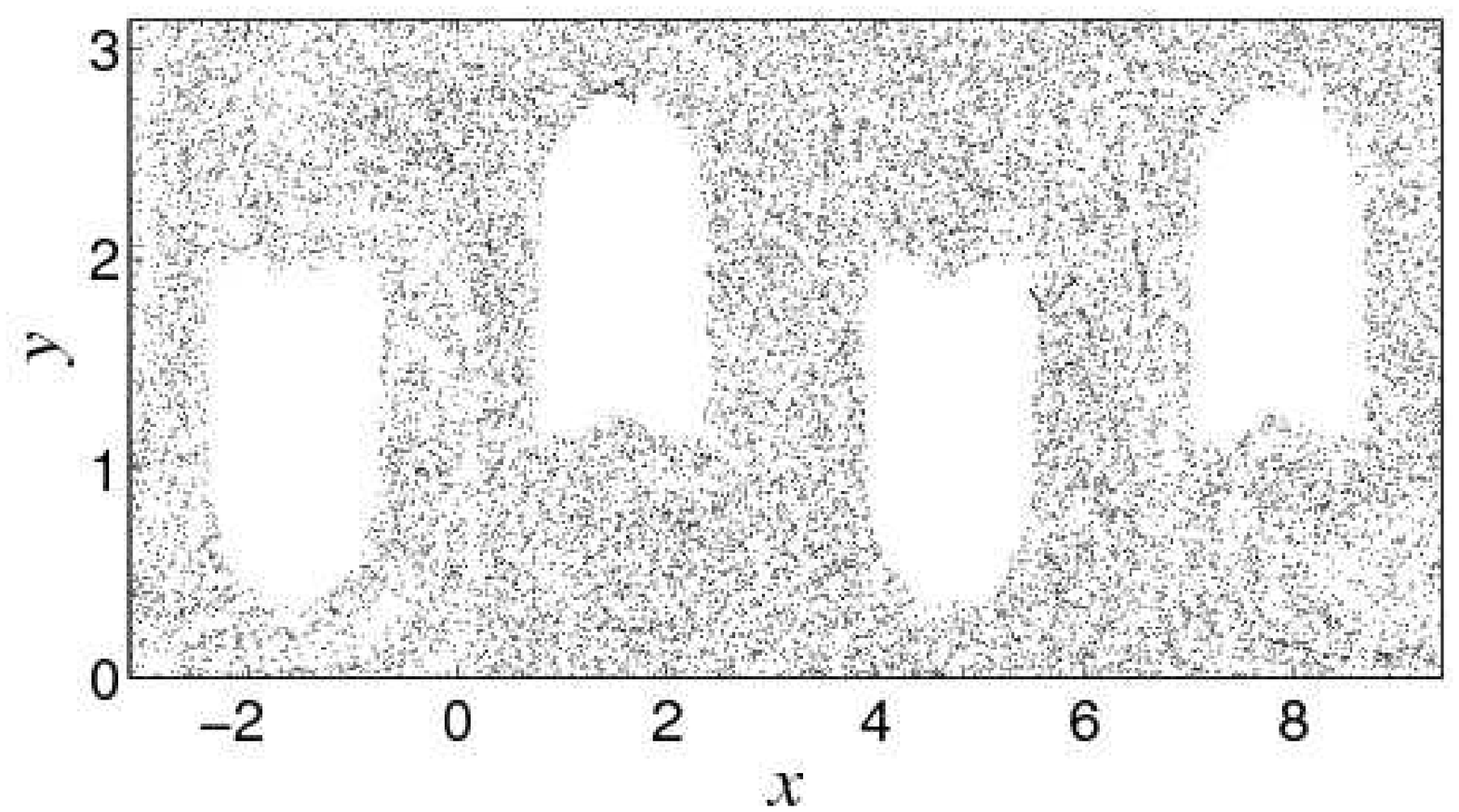}\includegraphics[%
  width=3cm]{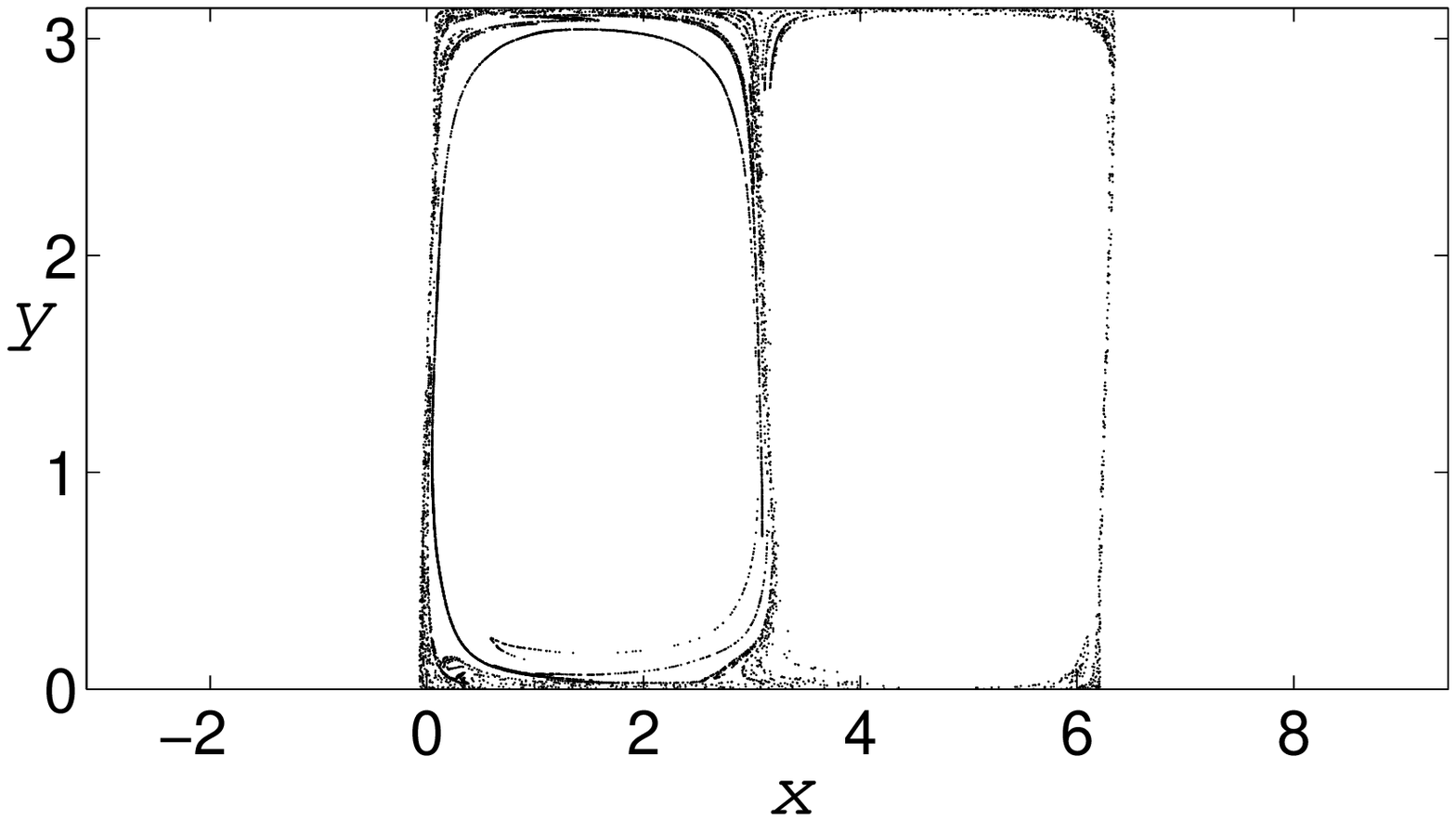}

\caption{\label{Fig3} Numerical simulation of the dynamics of a dye at $t=30$,
$t=50$, $t=70$ and $t=140$ (from top to bottom)~: left column
for the stream function~(\ref{1}) and right column for the
stream function~(\ref{streamcont}). The parameters are $\alpha=0.6$
and $\epsilon=0.63$.}
\end{figure}

In Fig.~\ref{Fig3}, we depict a numerical simulation of the dynamics of a dye in
the fluid. The left column shows the evolution of the tracers for
the stream function $\Psi_{\epsilon}$ given by Eq.~(\ref{1}). The
right column shows that the control term regularizes the dynamics and  prevents any spread of a dye within a cell which is limited by two
barriers created by the stream function $\Psi_{c}$ given by Eq.~(\ref{streamcont}).\\

For $\mid \alpha \epsilon\mid << 1$, the control term  (\ref{streamcont}) can be simplified such as:
\begin{eqnarray}
\label{31}
f_s(y,t)=-\frac{\alpha^2}{2} \sin 2y	\cos( \epsilon \sin t ) {C}_{\epsilon}(t).
\end{eqnarray}

We remark that the control term $f_s$ is of order $\alpha^2 \epsilon$ and does not depend on $x$ as it is expected from the method.\\
In order to test the robustness of the method and to try an experimentally
more tractable perturbation, we truncate the series given by Eq.~(\ref{serbessl}) by considering only the first term of the series
$C_{\epsilon}(t)$ in Eq.~(\ref{serbessl}) and $\cos( \epsilon \sin t )$ which leads to the following
stream function 
\begin{eqnarray}
\Psi_{s}(x,y,t)&=&\alpha\sin(x+\epsilon\sin t)\sin y \nonumber \\
 &&-{\alpha}^2 \mathcal{J}_{1}^{\epsilon}\sin 2y \cos  t.\label{psis}\end{eqnarray}
The perturbation has a norm  
 $\sup_{x,y,t} \mid V(x,y,t)\mid=$
  $\epsilon$ and the norm of the the simplified control term $f_s$ is 
\begin{eqnarray}
\label{16}
\sup_{y,t} \mid f_s(y,t)\mid={\alpha}^2
{\mathcal{J}_{1}^{\epsilon}} .	
\end{eqnarray}

Therefore the ratio $r$ between the norm of the control term and the perturbation is given by $r \approx {\alpha} \epsilon$.\\

\begin{figure}
  \begin{center}
    \epsfig{file=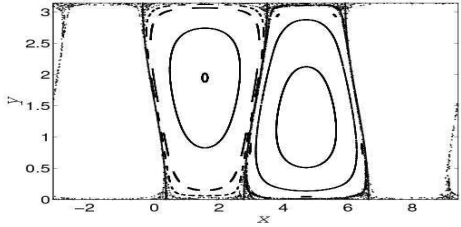,width=6.2cm,height=3cm}
 \caption{\label{Fig4}  Poincar\'{e} section of the stream function (\ref{psis}) . The parameters are $\alpha=0.6$ and $\epsilon=0.63$.}  
  \end{center}
\end{figure}

The dynamics of the stream function $\Psi_{s}$  is depicted in Fig.~\ref{Fig4} for $\epsilon=0.63$ and $\alpha=0.6$. We see that invariant surface has been created around $x=0$ (mod $2 \pi$) and that the control term 
still reduces significantly the chaotic advection but there are some advected particles along the channel. We observe that 
an other invariant surface has been created around $x=\pi$. This is due to the fact that the flow given by the stream function $\Psi_{s}$ is invariant under the symmetry $$ t\rightarrow t, x\rightarrow x+\pi, y\rightarrow y+\pi.$$ It results from this symmetry, a regularisation of the dynamics within the cell bounded by the two barriers  $x=0$ and $\pi$ (mod $2\pi$). This symmetry is an approximate one ( up to order $\mid \alpha^2 \epsilon \mid$) for the stream function $\Psi_{c}$. Thus the control term is able to regularize its dynamics also in this region.
 
\begin{figure}
\begin{center}
\epsfig{file=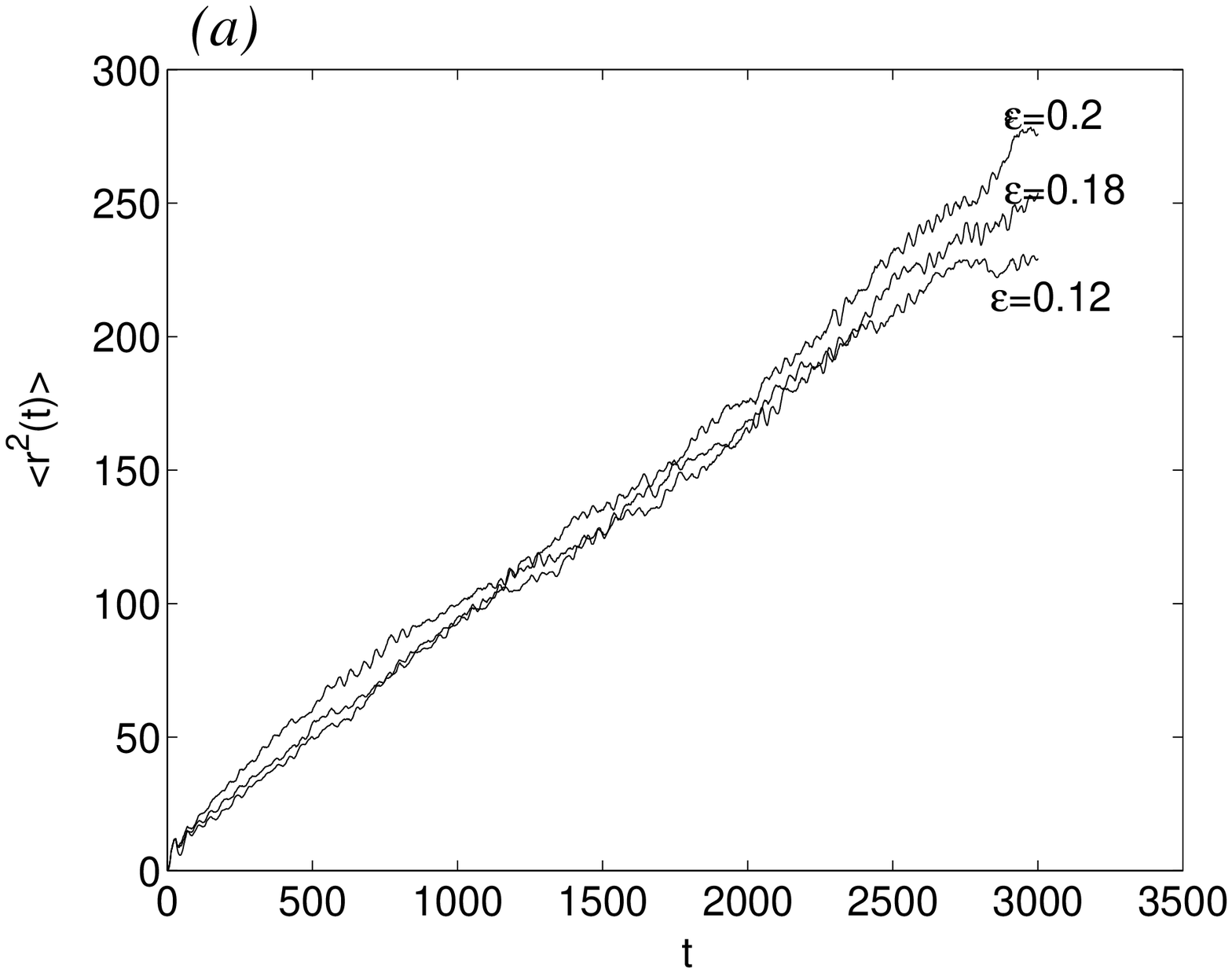,width=4cm,height=4cm} \epsfig{file=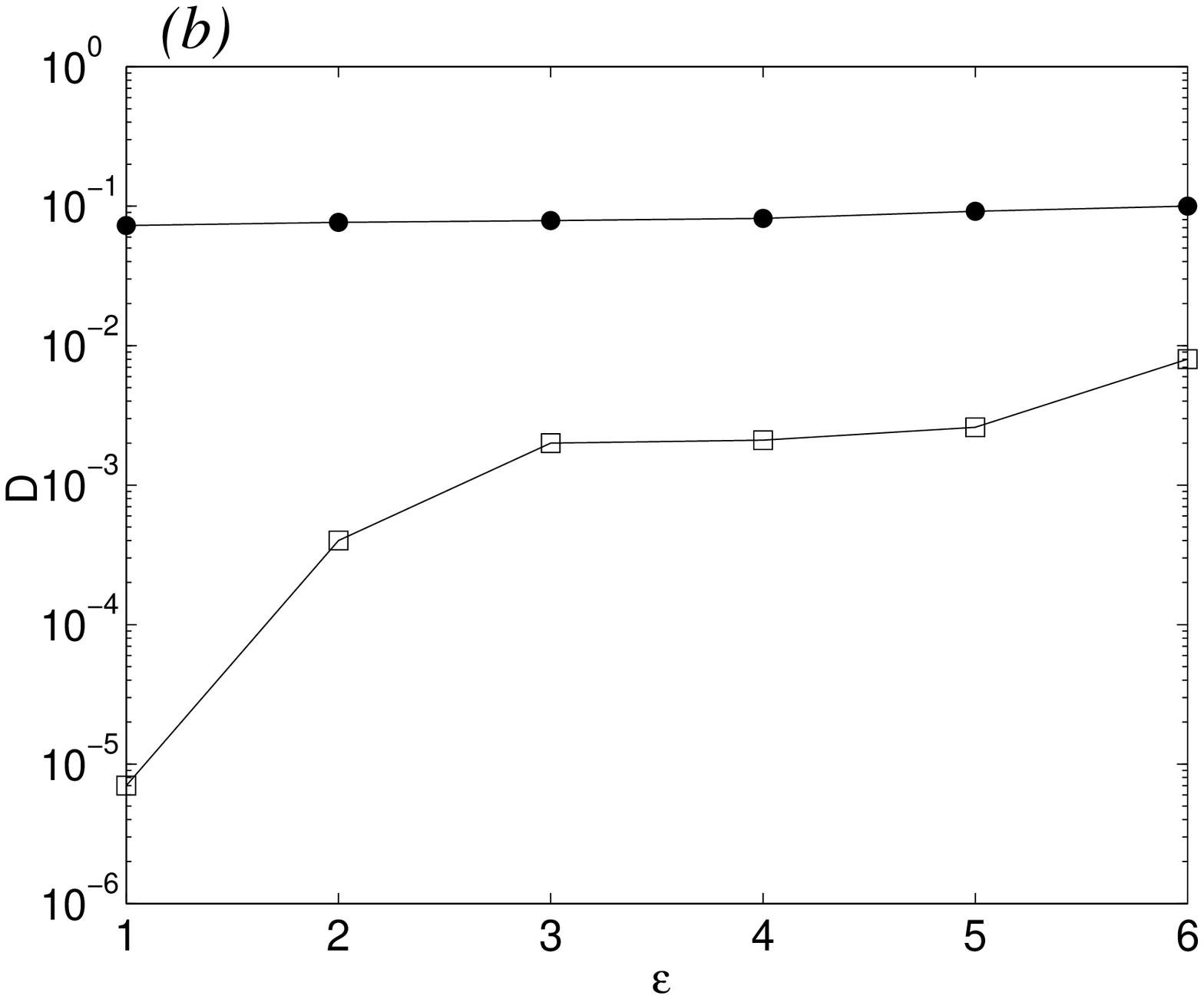,width=4cm,height=4cm}
\caption{\label{Fig5} $(a)$ Mean square dispacement $<r^2(t)>$ versus time $t$ obtained for Hamiltonian (\ref{1}) with three different values of $\epsilon=0.12, 0.18, 0.2$ and $(b)$ Diffusion coefficient D versus $\epsilon$ obtained for stream function (\ref{1}) (circle) and stream function (\ref{psis}) (square)}
\end{center}
\end{figure}

In order to see more clearly the effect of the control term, we study the diffusion properties of the system.
The mean square displacement $<r^2(t)>$ of a distribution of $\mathcal{M}$ particles (of order 3000) is computed as a function of time:
\begin{eqnarray}
	<r^2(t)>=\frac{1}{\mathcal{M}}\sum_{i=1}^{\mathcal{M}} \left\|{\bf x}_i(t)-{\bf x}_i(0)\right\|^{2},
\end{eqnarray}
where ${\bf x}_i(t)$, $i=1,...,\mathcal{M}$ is the position of the $i$-th particle at time $t$ obtained by integrating Hamilton's equations with initial conditions ${\bf x}_i(0)$. We find that $<r^2(t)>$ grows linearly  for the considered time interval  (see Fig.~\ref{Fig5}$(a))$. The transport is then assumed  to be described as normal diffusion and the corresponding diffusion coefficient can be determined from the slope of  $<r^2(t)>$ versus $t$:
\begin{eqnarray}
D=\lim_{t\rightarrow \infty} \frac{<r^2(t)>}{t}.
\end{eqnarray}
Figure ~\ref{Fig5}$(b)$ shows the values of $D$ as a function of $\epsilon$ with and without control term (\ref{psis}) determined from the mean square displacement for $ t > 1000$. We remark that the diffusion coefficient with the control term (\ref{psis}) is significantly smaller than the uncontrolled case.

\section{Conclusion}
We presented in this paper an application of a recently
developed control technique to Rayleigh-B\'enard convection.
We derived analytically the expansion of the control term required to reduce chaotic advection present
in the original problem. 
Using Poincar\'e sections, and tracking the dynamics of a dye, we showed the efficiency of the method. The slightly deformed streamlines due to the control terms prevent any large spread through the channel but confine the material lines inside some deformed rolls.
The chaotic behavior of the flow was significantly reduced by taking only the first order of control term. 
\bibliography{ifacsam}

H. Aref, {\it Stirring by chaotic advection}, J. Fluid Mech. \textbf{143}, 1 (1984).

S. Balasuriya, {\it Optimal perturbation for enhanced chaotic transport}, Physica D \textbf{202}, 155 (2005).

R.P. Behringer, S. Meyers and H. Swinney, {\it Chaos and mixing in geostrophic flow}, Phys. Fluids A \textbf{3}, 1243 (1991).

T. Benzekri, C. Chandre, X. Leoncini, R. Lima, M. Vittot, {\it Chaotic advection and targeted mixing},  Phys. Rev. Lett. \textbf{96}, 124503 (2006).

M.G. Brown, K.B. Smith, {\it Ocean stirring and chaotic low-order dynamics}, Phys. Fluids \textbf{3}, 1186 (1991).

R. Camassa, S. Wiggins {\it Chaotic advection in Rayleigh-B\'enard flow}, Phys. Rev. A \textbf{43}, 774 (1991).

C. Chandre, G. Ciraolo, F. Doveil, R. Lima, A. Macor, M. Vittot, {\it Channelling chaos by building barriers}, Phys. Rev. Lett. \textbf{94}, 074101 (2005).

J.M. Ottino, {\it The Kinematics of mixing: streching, chaos, and transport}, Cambridge U.P., Cambridge (1989).

T.H. Solomon, N.S. Miller, C.J.L Spohn, J.P. Moeur, {\it Lagrangian chaos: transport, coupling and phase separation}, AIP Conf. Proc. \textbf{676}, 195 (2003).

T.H. Solomon and I. Mezic {\it Uniform, resonant chaotic mixing in fluid flows}, Nature, \textbf{425}, 376 (2003). 

T.H. Solomon, J.P. Gollub, {\it Chaotic particle transport in Rayleigh-B\'enard convection}, Phys. Rev. A \textbf{38},6280 (1988).

A.D. Stroock, S.K.W. Dertinger, A. Ajdari, I. Mezic, H.A. Stone, G.M. Whitesides,{\it Chaotic mixer for microchannels}, Science \textbf{295}, 647 (2002).

M. Vittot, C. Chandre, G. Ciraolo and R. Lima, {\it Localized control for nonresonant Hamiltonian systems}, Nonlinearity \textbf{18},423 (2005).

H. Willaime, O.Cardoso, P. Tabeling, {\it Spatiotemporel intermittency in lines of vortices}, Phys. Rev. E, \textbf{48}, 288 (1993).

S. Wiggins, {\it Chaotic transport in dynamical systems}, Springer-Verlag (1992).

\end{document}